\PassOptionsToPackage{svgnames}{xcolor} %
\documentclass[10pt, conference, letterpaper]{IEEEtran}
\usepackage[english]{babel}
\usepackage{blindtext}
\usepackage{setspace}

\usepackage{amsmath,amsfonts} %
\usepackage{subcaption} %
\usepackage{algorithmic} %
\usepackage{algorithm} %
\usepackage{array} %
\usepackage{graphicx} 
\usepackage{booktabs} %
\usepackage{enumitem} %
\usepackage{hyperref} %
\usepackage[capitalize]{cleveref} %
\usepackage{physics} %
\usepackage{graphicx} %
\usepackage{textcomp} %
\usepackage{xcolor} %
\usepackage{tabularx} %
\usepackage{tikz} %

\usetikzlibrary{
  decorations.pathreplacing,
  automata,
  arrows,
  angles,
  quotes,
  calc,
  calligraphy,
  chains,
  decorations.pathmorphing,
  intersections,
  positioning,
  shapes}

\floatname{algorithm}{Protocol} %
\crefname{algorithm}{Protocol}{Protocols} %
\def\BibTeX{{\rm B\kern-.05em{\sc i\kern-.025em b}\kern-.08em T\kern-.1667em\lower.7ex\hbox{E}\kern-.125emX}}

\setlist[enumerate]{nolistsep,itemsep=3pt,topsep=3pt,leftmargin=*} %
\setlist[itemize]{nolistsep,itemsep=3pt,topsep=3pt,leftmargin=2em} %

\colorlet{royalred}{IndianRed}
\colorlet{royalgreen}{ForestGreen}
\colorlet{royalblue}{SkyBlue}
\colorlet{background}{yellow!60}
\begin{document}

\title{Control Flow Adaption: An Efficient Simulation Method For Noisy Quantum Networks}

\author{
  \IEEEauthorblockN{ %
  Huiping Lin\IEEEauthorrefmark{1}, %
  Ruixuan Deng\IEEEauthorrefmark{1}, %
  Chris Z. Yao\IEEEauthorrefmark{1}, %
  Zhengfeng Ji\IEEEauthorrefmark{1}\IEEEauthorrefmark{2}, %
  Mingsheng Ying\IEEEauthorrefmark{1}
  }
  \IEEEauthorblockA{
  \IEEEauthorrefmark{1}Tsinghua University\quad \IEEEauthorrefmark{2}Zhongguancun Laboratory\\
  Email: \href{lhp22@mails.tsinghua.edu.cn}{lhp22@mails.tsinghua.edu.cn},
  \href{dengrx24@mails.tsinghua.edu.cn}{dengrx24@mails.tsinghua.edu.cn},
  \href{chrisyaochrisyao@hotmail.com}{chrisyaochrisyao@hotmail.com},
  \\\href{jizhengfeng@tsinghua.edu.cn}{jizhengfeng@tsinghua.edu.cn},
  \href{yingmsh@mail.tsinghua.edu.cn}{yingmsh@mail.tsinghua.edu.cn}
  }
}

\maketitle 
\begin{abstract}
Quantum network research at both the software stack and hardware implementation level has become an exciting area of quantum information science.
Although demonstrations of small-scale quantum networks have emerged in the past decade, quantum communication and computation hardware remain scarce resources today.
As a result, the evaluation and validation of quantum network protocols primarily rely on classical simulators rather than real quantum networks.

This paper introduces a novel quantum network simulation method called control flow adaptation, which enhances standard tensor network simulations.
This method enables accurate and efficient simulations of many important quantum network protocols by carefully leveraging the control flow structures of them.
Furthermore, we have developed a prototype quantum network simulator, qns-3, as a module for ns-3.
This new module implements the control flow adaptation technique and applies it to a wide range of protocols.
All related code is open-sourced~\cite{GithubRepo}.

We believe the development of the control flow adaptation method and the qns-3 simulator represents a step forward in quantum network research, offering a versatile and scalable platform for testing and verifying quantum network protocols on classical computers.
\end{abstract}

\begin{IEEEkeywords}
  Quantum Networks, Network Simulation, Quantum Computing, Quantum Communication \end{IEEEkeywords}

\section{Introduction}
With the fast development of different quantum algorithms and their applications in different areas such as distributed computing~\cite{caleffi2018quantum,
  buhrman2003distributed,qiao2022quantum} and cryptography~\cite{elliott2003quantum, ma2008quantum,chaudhuri2015securing},
building quantum networks that offer quantum communication capabilities between different quantum computing units has become a vital task~\cite{sandilya2021quantum,kozlowski2019towards,wehner2018quantum,
  10.1145/3610251.3610557,10.1145/3610251.3610553,farahbakhsh2022opportunistic}.
In particular, methodologies and techniques developed for network research can be extended and adapted to facilitate the study of quantum networks significantly.

\textbf{Necessity of Quantum Network Simulation}: In classical network research, network simulation serves as software applications to evaluate computer network efficiency.
The advantages of network simulation are even more pronounced in the context of quantum networks.
Due to the increased complexity of communication networks, analytical methods fall short of understanding system behavior.
Quantum network simulation offers a secure environment to predict network performance without physically building an entire quantum network system,
significantly reducing hardware and maintenance costs as quantum equipment remains expensive today~\cite{de2021materials}.

\textbf{Challenges of Quantum Network Simulation}: The simulation techniques for classical networks cannot be simply transplanted into the quantum case due to the fundamental differences between classical and quantum networks:
\begin{enumerate}
  \item Quantum information is ``continuous'' as the amplitudes of a qubit could be any complex number, making it impractical to represent them using classical bits.
  Therefore, the simulation of quantum networks cannot be achieved directly using classical network simulators.
  \item Quantum features such as entanglement and superposition~\cite{nielsen2001quantum,vedral2014quantum,watrous2018theory}
  adds significant simulation complexity.
  Quantum systems with $n$ qubits require $2^n$ complex numbers to represent their state.
  As the number of qubits increases, the computational resources needed to store and manipulate these states grow exponentially.
  \item Quantum systems are highly sensitive to their environment, leading to decoherence and noise~\cite{clerk2010introduction}.
  Accurately modeling quantum noise is challenging but essential for realistic simulations~\cite{wehner2018quantum}.
\end{enumerate} 

\textbf{Major Innovations:} To tackle these challenges, our work encompasses two key innovations. First, we introduce a novel technique called Control Flow Adaptation (CFA), designed to enhance simulation efficiency across various quantum protocols. 
Second, we develop a quantum simulator, qns-3, as a module for ns-3~\cite{riley2010ns}, integrating the CFA technique within this simulator.

In quantum networks, many protocols can be modeled in the LOCC (local operations and classical communication) scheme.
In these protocols, classical information received by different parties influences future quantum operations, leading to the required construction of a complex backend to track quantum states if done straightforwardly.
For example, the tensor network method is a powerful technique for the classical simulation of quantum information, but as the protocols become more intricate and the number of qubits increases, the contraction complexity of the tensor networks often grows exponentially.
The need to simulate the noise behavior in quantum networks marks a further challenge for tensor network methods as noises add many new contractions in the tensor network representation (see e.g.\ \cref{fig:chain-swapping}).
Additionally, quantum measurements, which bring randomness in the simulation of quantum network protocols, are another source of complications.
To obtain an average simulation result when verifying a protocol, one may need to employ a Monte Carlo method and run the simulation numerous times.

The CFA method we propose in this context performs a control flow analysis to determine the necessary information required for future rounds of the LOCC protocols.
By compressing and adapting the classical control flow in a way that aligns with the capabilities of tensor networks, we can efficiently handle the complexities of many quantum network protocols.
This careful adaptation ensures that the control flow information integrates seamlessly into the tensor network simulation framework.
Instead of faithfully simulating all quantum measurements and revealing one possible branch, this method provides a mechanism that is able to integrate all possible results in a single round.

This paper also presents the first quantum network simulator designed based on ns-3 with suitable APIs respecting ns-3's convention~\cite{henderson2008network}, named as \emph{qns-3}.
Given ns-3's position as the most widely utilized network simulator for classical network research, we are confident that our work will provide valuable assistance to scholars and engineers with a focus on classical networks who are interested in delving into the realm of quantum network research.
Currently, qns-3 supports a tensor-network backend enhanced by the CFA method.
We plan to expand the capabilities of qns-3 in the near future.

\textbf{Framework of the Simulator:} Our simulator leverages the discrete-event simulation model of ns-3, incorporates the management of quantum operations,
quantum communications, and quantum noises within its event scheduler, and implements the control flow adaption method.
Specifically, certain types of noises are time-dependent and are applied when required, considering the elapsed time between events.
The data structure we choose to model all the quantum ingredients is tensor networks, which can effectively represent all the essential components needed for a quantum network, including quantum states, gates, and noises.

\begin{figure}[ht]
  \centering \includegraphics[scale=0.6]{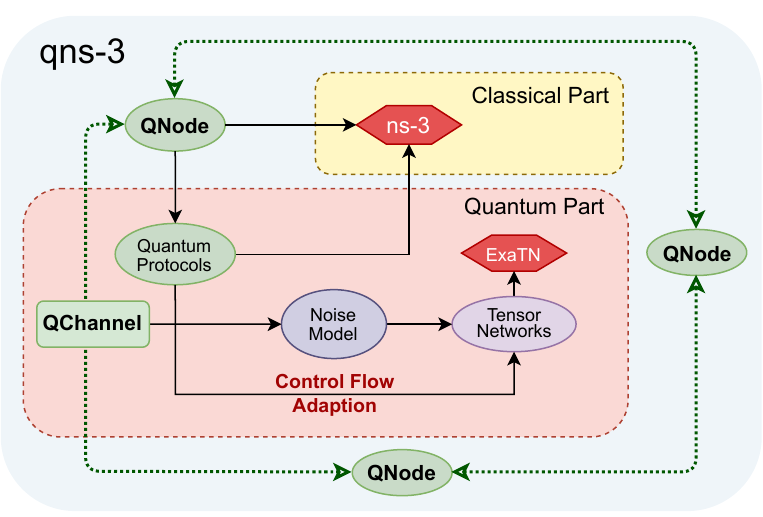}

  \caption{Framework of the qns-3 simulator.
  The network model consists of green modules representing quantum nodes (\textsf{QNode}) that communicate (green dotted lines) using quantum channels (\textsf{QChannel}) and various quantum protocols.
  The simulator utilizes the \textsf{ns-3} framework for classical communication, adhering to its standards.
  Quantum operations are simulated using tensor networks through \textsf{ExaTN}~\cite{lyakh2022exatn}.
  Additionally, the control flow adaption is adopted for simulating quantum protocols, as discussed in \cref{sec:method}.
  Further details on the simulator structure are provided \cref{sec:structure}.}\label{fig:intro}
\end{figure}

The framework of qns-3 is shown in \cref{fig:intro}, where the basic components of the simulator and the network model are illustrated.
The network contains two main classes of components: quantum nodes (\textsf{QNode}) and quantum channels (\textsf{QChannel}).
Quantum nodes model hosts in a quantum network that can run different quantum programs or algorithms locally.
They are also equipped with classical and quantum network devices providing communication capabilities.
Quantum channels refer to physical pipelines that can transmit quantum information and are mathematically modeled as completely positive trace-preserving maps~\cite{nielsen2001quantum} in quantum information theory.
In our setup, quantum channels are used for the entanglement generation or direct qubit transportation between quantum nodes.

\setlength{\belowcaptionskip}{-15pt}

\textbf{Organization of the Paper}: \Cref{sec:related} reviews the related existing works.
\cref{sec:pre} exhibits some of the necessary preliminaries.
\cref{sec:method} provides an exhaustive description of our key technical:
control flow adaptation.
\cref{sec:netowrk-model} gives an overview of the quantum network model incorporated in qns-3, and 
\cref{sec:structure} provides an in-depth examination of our implementation of qns-3.
\cref{sec:experiment} presents the experimental results of our simulator.
Finally, \cref{sec:conclusion} gives a concise conclusion.

\begin{table*}[ht]
  \centering \begin{tabular}{l l l c c c c c}
  \toprule Year & Simulator & Language & Open-Source & Generality & Discrete-Event & Faithful Simulation & Free of MC\\
  \midrule 2023 & QNS-3  & C++ & Y & Y & Y & Y & Y\\
  \midrule 2022 & QNET & Python & Y & Y & N & Y & N\\
  2020 & NetSquid & Python & N & Y & Y & Y & N\\
  2020 & SeQUeNCe & Python & Y & Y & Y & Y & N\\
  2020 & QuNetSim & Python & Y & Y & N & Y & N\\
  2019 & QuISP & C++ & Y & Y & Y & N & Y\\
  2021 & QuNet & Julia & Y & N & N & Y & N\\
  2015 & qkdX & C++ & Y & N & Y & Y & N\\
  \bottomrule  \end{tabular}
 \caption{Comparison of different simulators.
 ``Y'' means the feature is supported and ``N'' means not.
 ``Free of MC'' in the last column means whether a Monte Carlo simulation can be avoided to obtain an average result of a quantum protocol.
 }\label{table:simulators}
\end{table*}

\section{Related Works}\label{sec:related}

We review the existing works that have been developed in the past few years for building quantum network simulators.
Currently, various quantum network simulators have been released, including general simulators such as 
QNET~\cite{fang2022quantum}, QuISP~\cite{satoh2022quisp},
QuNetSim~\cite{diadamo2021qunetsim}, SeQUeNCe\cite{wu2021sequence},
SQUANCH~\cite{bartlett2018distributed}, and NetSquid~\cite{coopmans2021netsquid}, as well as specialized simulators such as qkdX~\cite{mailloux2015model} and QuNet~\cite{leone2021qunet}.
We briefly overview some of these simulators and compare them to qns-3.

\textit{NetSquid} is a software tool designed to model and simulate scalable quantum networks, building a comprehensive framework that includes physical model composition, quantum and classical communication ports, and all necessary hardware components from scratch.
It integrates various essential technologies, such as a discrete-event simulation engine, a specialized quantum computing library, a modular framework for modeling quantum hardware devices, and an asynchronous programming framework for describing quantum protocols.
As a result, \textit{NetSquid} stands out as one of the most comprehensive and well-designed network simulators currently available.
However, it is important to note that \textit{NetSquid} is not open-source.
\textit{NetSquid} faithfully simulates all events in a network.
But as far as we know, no specific optimization is done for network protocols.

\textit{QuISP} is also an event-driven simulator.
It is designed to simulate large-scale quantum networks by tracking only 7
specific error types using a vector for each qubit.
The error tracking idea allows it to simulate large networks, but the method of solely recording specific errors fails to faithfully model the behavior of the network, rendering it inappropriate and inaccurate in many situations.

In \cref{table:simulators}, we list many other quantum network simulators and compare them to qns-3.
We can see that most simulators are typically open-sourced and are primarily developed using Python, with two noticeable exceptions, QuISP and qkdX, which are implemented in C++.
These simulators are built upon OMNeT++, a component-based C++ simulation framework.
Notably, ns-3, the very popular discrete-event network simulator for the Internet, is not currently supported by any existing quantum simulators.
Moreover, most simulators ensure accurate simulation by faithfully replicating all actions and tracking the states of every qubit in the network without any time or space optimization.
In contrast, qns-3 is built upon ns-3, making it potentially more accessible to researchers and engineers familiar with traditional networks.
It focuses on simulation efficiency, enabling large-scale simulations that are valuable for network research.

\section{Preliminaries}\label{sec:pre}

In this section, we exhibit some necessary definitions and notations for readers unfamiliar with quantum computation.
We follow the standard notations in~\cite{nielsen2001quantum} throughout the paper.

Quantum networks encode information into quantum states, and the most commonly used quantum system is a \textit{qubit}.
It has two basis states $\ket{0}$ and $\ket{1}$, analogous to $0$ and $1$ in a binary bit in classical computing.
Unlike a classical bit that is either a $0$ or a $1$, the qubit can be in a superposition of the two base states, written as $\ket\psi = \alpha\ket{0}+\beta\ket{1}$, where $\alpha$ and $\beta$ are complex probability amplitudes satisfying $\norm{\alpha}^2+\norm{\beta}^2=1$.
$\ket{0}$ and $\ket{1}$ can also be seen as the basis of a $2$-dimensional Hilbert space $\mathcal H$, and thus $\ket{\psi}$ can also be seen as a column vector in $\mathcal H$. $\bra{\psi}$ denotes the Hermitian conjugate of $\ket{\psi}$.

The notion of \textit{entanglement} describes a special phenomenon in multi-qubit states where the qubits of the system cannot be described individually independent of the state of the others.
Entangled states are valuable resources for carrying out many quantum network protocols~\cite{panigrahy2022optimal,kolar2022adaptive,zeng2022multi}.
The most widely used entanglement states are the four EPR states, written as 
\begin{align*}
  \ket{\phi_{\pm}} = \frac{1}{\sqrt 2}\left(\ket{00} \pm \ket{11}\right),\;
  \ket{\psi_\pm} = \frac{1}{\sqrt 2}\left(\ket{01} \pm \ket{10}\right).
\end{align*}
Without loss of generality, when mentioning EPR pairs, we refer to $\ket{\phi_+}$ by default.

\textit{Bell measurement} denotes a measurement of two qubits $q_1,q_2$ that determines which of the four EPR states the two qubits are in, and is achieved by applying a \textsf{CNOT} gate on $q_1$ and $q_2$, a Hadamard gate on $q_1$,
and then measuring the two qubits using the standard basis.
It is widely used in quantum communication.

A \textit{pure state} is a state which can be described by a single vector.
All quantum states presented above are pure states.
A \textit{mixed state} is a statistical ensemble of pure states, described by a density matrix: $\rho = \sum_i p_i\ket{\psi_i}\bra{\psi_i}$, with $\sum_i p_i=1$.

\textit{Fidelity} is a measure of the ``closeness'' of two quantum states.
For any two mixed states $\rho$ and $\sigma$, their fidelity is expressed as:
\vspace*{-2mm}
\begin{equation*}
  F(\rho, \sigma)={\Bigl( \tr \sqrt{\sqrt{\rho} \sigma \sqrt{\rho}} \Bigr)}^2.
\end{equation*}

In the following paragraph, we will mainly consider the fidelity between a mixed state $\rho$ and a pure state $\ket{\psi}$.
In this case, the fidelity is expressed as $F(\rho, \ket{\psi})= \tr (\bra{\psi}\rho\ket{\psi})$.

The \textit{no-cloning theorem}~\cite{nielsen2001quantum,lindblad1999general}
states that it is impossible to create a copy of an unknown quantum state.
It is essential for the understanding of quantum information and explains many design choices in quantum networks.

\textit{Local operations and classical communication} (LOCC) refers to a class of operations that can be performed separately on individual quantum subsystems,
followed by classical communication between the parties involved.
LOCC holds significant importance in quantum network research as the majority of quantum network protocols are LOCC-based, complemented with EPR distribution.
A general description of network protocols that can be viewed as a multi-round LOCC protocol is given in \cref{protocol:LOCC}.

\textit{Tensor Networks} refer to a data structure and a related collection of important techniques used to understand and interpret multi-linear maps.
A tensor network is a collection of tensors (multi-dimensional arrays) that are interconnected, representing complex linear algebraic computations graphically.
Each tensor is a generalization of vectors and matrices to higher dimensions,
with the connections between them representing summations over shared indices.
It is perhaps the most successful numerical simulation method for quantum computation and can easily represent quantum states, measurements, and channels commonly encountered in quantum information.
We refer the reader to~\cite{orus2019tensor,biamonte2020lectures} for an in-depth discussion on tensor networks.

Quantum systems are delicate and easily suffer from noises.
In classical computing, because binary bits only have two ``states'', we only need to consider bit flip errors and loss errors.
Quantum noise is completely different from classical noise and is modeled as completely positive trace-preserving (CPTP) maps between spaces of operators.
Their effects can be represented by a set of Kraus operators $\{E_i\}$:
\begin{equation*}
  \rho \mapsto \sum_i E_i\rho E_i^\dagger,
  \text{ where } \sum_i E_i^\dagger E_i = I.
\end{equation*}

In quantum networks, quantum noise is ubiquitous.
Even when no operations happen, quantum noise gradually introduces decoherence in the quantum system.
Hence, modeling the quantum noises in quantum networks is tricky, and we will present a reasonably simplified noise model in \cref{sec:structure}.

\section{Key Technique}\label{sec:method}

In this section, we introduce the key technical method we use to improve the efficiency of tensor network-based simulation for LOCC protocols: control flow adaption.
We will give an overall description of the method first and discuss how to apply it in two motivating examples.
Finally, we highlight the advantages this method brings.
The experimental results in Section~\ref{sec:experiment} confirm the advantages numerically.

\subsection{General Description}
\begin{algorithm}[ht]
  \floatname{algorithm}{Protocol}
  \caption{$L$-round LOCC}\label{protocol:LOCC}
  \begin{flushleft}
  \textit{Parties:} $P_1, P_2, \ldots, P_n$.

  \textit{The protocol:} For $i=1, \ldots, L$:
  \end{flushleft}
  \begin{enumerate}
  \item \textbf{Local Operation:} $P_1,P_2,\ldots,P_n$ apply local operations on their local qubits conditioned on all messages they received in the previous round $c_1^{i-1},\ldots,c_n^{i-1}$ (if $i=1$, $c^{0}_{j}$ are empty strings), and obtain new classical information $c_1^{i},\ldots,c_n^{i}\in\{0,1\}^{*}$ that they want to communicate.
  \item \textbf{Classical Communication:} $P_1,P_2,\ldots,P_n$ distribute $c_1^i,\ldots,c_n^i$ classically.
  \end{enumerate}
\end{algorithm}

\setlength{\belowcaptionskip}{-6pt}
\begin{figure*}[htb]
  \centering \begin{subfigure}[b]{0.85\textwidth}
  \includegraphics[width=1\linewidth]{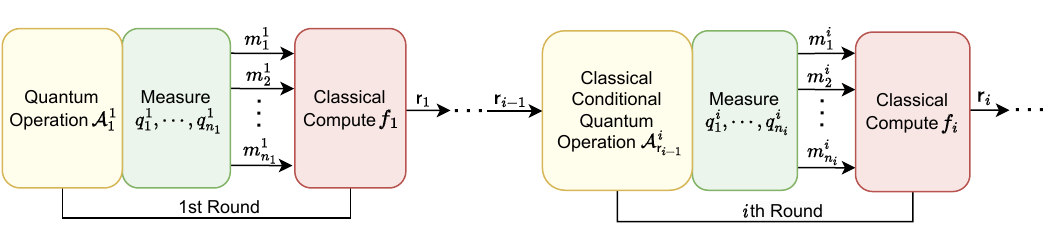} 
  \caption{LOCC protocol after analyzing control flows.}\label{fig:Tech1}
  \end{subfigure}
  
  \begin{subfigure}[b]{0.85\textwidth}
  \includegraphics[width=1\linewidth]{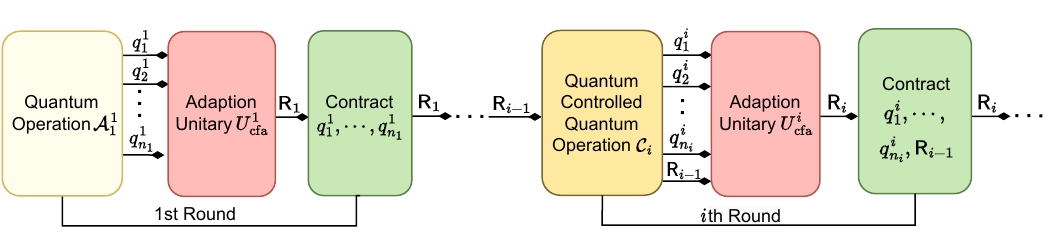} 
  \caption{Simulate LOCC protocol with adaptation unitary and deferred measurement.}\label{fig:Tech2}
  \end{subfigure}
  
  \caption{$L$-round LOCC protocol with control flow adaption.
  \cref{fig:Tech1} describes the protocol after analyzing control flows,
  and \cref{fig:Tech2} describes how to use control flow adaption to simulate the protocol.
  Despite the first round, the classical computation process is replaced as an adaptation unitary in each round (red modules, conditioned by normal arrows), and all classical controlled operations are replaced as quantum controlled operations (background modules, controlled by diamond-shaped arrows).
  All the measurements are deferred to the end of the next round by tensor contraction (green modules).
  The modules with deeper colors correspond to the modules with the same lighter colors changed by adaptation unitary and deferred measurement.}\label{fig:tech}
  
\end{figure*}
  
In each round of an LOCC protocol, classical messages will decide some actions in the next round.
In a quantum protocol, these classical messages come from either classical input or quantum measurements.
When the parties communicate classically, they may exchange their measurement outcomes to determine their follow-up actions.
By analyzing control flows of the protocols in each round, conditioned on the messages $\mathsf r\in\{1,2, \cdots, k \}$, their actions $A_{\mathsf r}\in\{A_1,A_2,\cdots, A_k\}$ are described as quantum operations ${\mathcal A}_{\mathsf r}\in\{{\mathcal A}_1, {{\mathcal A}_2}, \cdots ,{\mathcal A}_k\}$.
\footnote{For example, if there are four parties $P_1,\ldots,P_4$ each holding a qubit, and an action $A=$~``$P_1$ applies $X$ gate; $P_2$ applies $Y$ gate;
  $P_3$ applies $Z$ gate; $P_4$ does nothing'', then ${\mathcal A}(\cdot) = X_1\otimes Y_2\otimes Z_3 (\cdot) X_1^\dagger\otimes Y_2^\dagger\otimes Z_3^\dagger$.}

If the measurement outcomes determine further actions, faithfully simulating the measurements (which is achieved by generating a random value and determining the measurement result accordingly) and using the outcomes as classical control flow only allow us to work on one branch of multiple possibilities.
Then, one may need to follow a Monte Carlo method and run the simulation numerous times to obtain an average simulation result.
Our simulator supports such straightforward simulations but also provides a method that enables users to represent the quantities of interest exactly in a tensor network.

We first informally introduce the intuition behind our method.
Each round of a network protocol is modeled by a set of quantum operations ${\mathcal A}_{\mathsf r}$ determined by the information in a classical register ${\mathsf r}$.
A straightforward simulation is to calculate ${\mathsf r}$ and apply the corresponding quantum operation.
If the classical value of ${\mathsf r}$ has $k$ different possibilities, the direct simulation method can only investigate only one of the $k$ possible branches.
Our idea is to use a quantum register $\textsf R$ containing a superposition state of $\{\ket 1, \ket 2, \ldots, \ket k\}$ to store the content in ${\mathsf r}$, and change the quantum operations conditioned on ${\mathsf r}$ to a quantum control operation controlled by $\textsf R$ using the delayed measurement principle~\cite{nielsen2001quantum}.
More details are provided in the following discussions.

\subsection{Control Flow Adaption}
We now formally describe the control flow adaption method, which consists of three stages: \textit{control flow compression, adaptation unitary design, and message qubit contraction.}
For any $L$-round LOCC protocol, let's consider the $i$-th ($i=1, \ldots, L$)
round of the protocol.
The method works as follows:

\subsubsection{Control Flow Compression}

The first stage is to analyze the information extracted from measurements.
The $i$-th round receives from the previous round a register of qubits named $\textsf R_{i-1}$, holding the necessary control information needed for the current round.\footnote{If $i=1$, there's no previous control flow and we take $\mathsf{R}_{0}$ to be an empty register.}
The parties as a collection apply the action $A^i_{j} \in \{A^i_1,\ldots, A^i_{k_{i-1}}\}$ conditioned on $j\in\{1,2,\ldots,k_{i-1}\}$.
The actions are described as quantum operations $\{{\mathcal A}^i_1, \ldots, {\mathcal A}^i_{k_{i-1}}\}$, acting on all qubits represented in the tensor network.
So we can design a quantum operation ${\mathcal C}_{i}$ satisfying ${\mathcal C}_{i}(\ket{j}\bra{j} \otimes \rho^S) =
\ket{j}\bra{j} \otimes {\mathcal A}^{i}_{j} (\rho^S)$,
with $\rho^S$ denoting the state of the whole system.
It's easily verified that if each ${\mathcal A}^{i}_{j}$ has the Kraus operator set $\{E^{i,k}_{j} \}$, then the Kraus operator set $ \{\tilde E^{i,k}_{j} :=\ket{j}\bra{j}\otimes E^{i,k}_{j} \}$ describes quantum operation ${\mathcal C}_{i}$.

The parties measure qubits $q_1^{i},\ldots,q_{n_i}^{i}$ and let the measurement outcomes be $m^{i}_1, \ldots, m^{i}_{n_{i}}$ \footnote{For simplicity, we focus on the situation that measurements do not depend on $\mathsf{R}_{i-1}$.
The dependence can be solved by using a controlled swap to move the qubits.}.
Then they compute a function $f_i$, which takes the measurement outcomes $m^{i}_1,\ldots,m^{i}_{n_{i}}$ and previous control information $j$ as inputs, and outputs $j'\in\{1,2,\ldots,k_i\}$,
the classical control flow for the next round.
The role of the function $f_{i}$ is to partition its input space by the control pattern of each input.

\subsubsection{Adaption Unitary Design}

The second stage is to design a quantum adaptation unitary $U^i_{\rm cfa}$,
which simulates the computation of the function $f_i$.
$U^i_{\rm cfa}$ acts on qubits $q^{i}_1,\ldots,q^{i}_{n_{i}}$, $\textsf R_{i-1}$
and some ancillary qubits $\textsf R_{i}$ that are the register for the next round and initialized to $\ket{0}$, satisfying \begin{align*}
  & U^i_{\rm cfa} (|m^{i}_1, \ldots, m^{i}_{n_{i}}, j\rangle \otimes \ket{0}_{\textsf R_{i}} )\\
  =\; & (U^\prime|m^{i}_1, \ldots, m^{i}_{n_{i}}, j\rangle) \otimes \ket{j'}_{\textsf R_{i}}
\end{align*}
for all possible outcomes $m^{i}_1, \ldots, m^{i}_{n_{i}}$ and previous control information $j$. Here, the unitary $U'$ could be arbitrary as the qubits it acts on will be traced out anyway.
This freedom is sometimes critical to have efficient simulations (see \cref{sec:examples} for an example).
The basic idea is to choose a particular $U^{i}_{\rm cfa}$ so that the resulting tensor network is efficient to contract.

\subsubsection{Message Qubit Contraction}

The last step is a simple one.
After the classical control flow compression and adaption, we know how to extract the compressed information $\textsf{R}_{i}$ that determines the actions of future rounds.
This means that the message qubits storing $m^{i}_{1}, \ldots, m^{i}_{n_{i}}$ are no longer needed and can be safely discarded.
This corresponds to taking a partial trace of qubits $q^{i}_1,\ldots,q^{i}_{n_{i}}, \textsf R_{i-1}$ and can be implemented by simple tensor contractions in the tensor network.
This step potentially saves the memory usage of the simulation algorithm.

\subsection{Examples}\label{sec:examples}
In the following, we present the control flow adaption analysis for the chained entanglement swapping and nested distillation protocols to illustrate its use.
\subsubsection{Chained Entanglement Swapping}
Consider first the chained entanglement swapping protocol in \cref{pro:chain-swapping}.
In the protocol, the parties run a series of entanglement swapping simultaneously and establish an EPR pair between the first and the last participants.

\setlength{\textfloatsep}{5pt}
\begin{algorithm}[h]
  \floatname{algorithm}{Protocol}
  \caption{Chained Entanglement Swapping}\label{pro:chain-swapping}
  \begin{flushleft}
  \textit{Sender:} $A_1$, \textit{Receiver:} $A_n$, \textit{Routers:} $A_2,\ldots,A_{n-1}$.\\
  \textit{Goal:} $A_{i}$ and $A_{i+1}$ have the ability to share EPR pairs for $i=1,2,\ldots, n-1$. The final goal is to obtain a shared EPR pair between $A_1$ and $A_n$.\\
  
  \textit{The protocol:}
\end{flushleft}
  \begin{enumerate}
  \item \textbf{EPR Generation.}
  Generate shared EPR pairs between $\langle A_i,A_{i+1}\rangle$ for all $i=1,\ldots,n-1$.
  \item \textbf{Local Operation.}
  All parties except for $A_1$ and $A_n$ apply Bell measurement on their two qubits $q^1_{i},q^{2}_i$, and obtain outcomes $(m^1_{i},m^{2}_i)\in\{0,1\}^2$.
  \item \textbf{Classical Communication.}
  All parties use classical communication to inform $A_n$ of the measurement outcomes: $(m^1_{i},m^{2}_i)$ for all $i=2,\ldots,n-1$.
  \item \textbf{Local Operation.}
  $A_n$ applies local corrections $X^{\oplus_i m^1_i}$ and $Z^{\oplus_i m^2_i}$ according to measurement outcomes.
  \end{enumerate}
\end{algorithm}

We follow the three-step recipe of control flow adaption to analyze this protocol.
First, in the control flow compression step, it is easy to come up with function $f$, which computes the two parity bits of measurement outcomes 
\begin{equation*}
  f(m_2^1, \ldots, m_{n-1}^1, m^2_2, \ldots, m_{n-1}^2)
  = \Bigl( \bigoplus_i m^1_i, \bigoplus_i m^2_i \Bigr),
\end{equation*}
as all we need in the second round are these two parity bits.
Here, $(\oplus_i m^1_i,\oplus_i m^2_i) \in\{0,1\}^2$ could be represented by only $\textsf R^1,\textsf R^2$.

Next, we need to design the adaption unitary $U_{\rm cfa}$ that implements the function $f$.
We focus on the $X$ control information $r^1=\oplus_i m^{1}_{i}$ first.
There are two simple ways to define a unitary that computes $\oplus_i m^1_i$;
both methods use the \textsf{CNOT} gate $n-2$ times.
In the first, for $i=2,\ldots,n-1$, the control qubit is $q^{1}_{i}$ and the common target qubit is $\textsf R^1$.
In the second, for $i=2,\ldots,n-2$, the control qubit is $q^{1}_{i}$ and the target qubit is $q^{1}_{i+1}$; and finally, the control qubit is $q^{1}_{n-1}$
is and the target qubit is $\textsf R^1$.

The first comes from a straightforward delayed measurement technique for computing the parity.
The effect of this choice of $U_{\rm cfa}$ on the tensor network is shown in \cref{fig:chain-swapping}.
The tensor diagram shows that this choice is problematic.
As the number of parties $n$ grows, the tensor network will become complicated and the tensor contraction complexity will grow exponentially.
This can be seen by considering the tensor nodes for $A_{n-1}$, which contains order $n$ outgoing links connecting to tensor nodes in $A_{i}$'s for $i=1, 2, \ldots, n-2$.
So no matter where we put the tensor nodes for $A_{n-1}$ in the contraction sequence, either before or after we add these nodes, there will be order $n$ open links.

As illustrated in \cref{fig:chain-swapping-adapt}, the second design of $U_{\rm cfa}$ leads to tensor nodes arranged in a chain, which is much easier to handle in tensor network contraction algorithms.
We can arrange the tensor nodes for $A_{1}, A_{2}, \ldots, A_{n}$ sequentially,
and the contraction complexity will be linear in $n$.
So by choosing a suitable adaption unitary, we can adapt the control flow implementation so that it works more smoothly with tensor network techniques.

\def\eprsep{1.4}
\def\linksep{4.5}

\begin{figure}[htb]
  \centering \begin{tikzpicture}[scale=.25,
  empty/.style={circle,inner sep=0,outer sep=0,minimum width=1.2mm}, %
  error/.style={draw,circle,fill=royalred,inner sep=0,outer sep=0,minimum width=1.2mm}, %
  gate/.style={draw,circle,fill=royalblue,inner sep=0,outer sep=0,minimum width=1.2mm}, %
  adapt/.style={draw,circle,fill=royalgreen,inner sep=0,outer sep=0,minimum width=1.2mm}, %
  leg/.style={inner sep=0,outer sep=0}, %
  link/.style={rounded corners=1mm}]

  \coordinate (O) at (0,0);
  \foreach \i in {1,...,3} {
  \foreach \j in {1,...,7} {
  \node[empty] (U\i\j) at (.8*\i,\linksep*\j) {};
  \node[empty] (L\i\j) at (.8*\i,\linksep*\j-\eprsep) {};
  \node[empty] (V\i\j) at (-.8*\i,\linksep*\j) {};
  \node[empty] (S\i\j) at (-.8*\i,\linksep*\j-\eprsep) {};

  }
  }

  \draw[fill=background!30,background!30] (-16,\linksep - 0.5*\eprsep - 3) rectangle (13.5,\linksep - 0.5*\eprsep);
  \draw[fill=background!10,background!10] (-16,\linksep - 0.5*\eprsep) rectangle (13.5,2*\linksep - 0.5*\eprsep);
  \draw[fill=background!30,background!30] (-16,2*\linksep - 0.5*\eprsep) rectangle (13.5,3*\linksep - 0.5*\eprsep);
  \draw[fill=background!10,background!10] (-16,3*\linksep - 0.5*\eprsep) rectangle (13.5,4*\linksep - 0.5*\eprsep);
  \draw[fill=background!30,background!30] (-16,4*\linksep - 0.5*\eprsep) rectangle (13.5,5*\linksep - 0.5*\eprsep);
  \draw[fill=background!10,background!10] (-16,5*\linksep - 0.5*\eprsep) rectangle (13.5,6*\linksep - 0.5*\eprsep);
  \draw[fill=background!30,background!30] (-16,6*\linksep - 0.5*\eprsep) rectangle (13.5,7*\linksep - 0.5*\eprsep);
  \draw[fill=background!10,background!10] (-16,7*\linksep - 0.5*\eprsep) rectangle (13.5,7*\linksep - 0.5*\eprsep + 2.5);

  \node at (-14.5, \linksep-0.5*\eprsep-1.5) {$A_{8}$};
  \node at (-14.5, 1.5*\linksep-0.5*\eprsep) {$A_{7}$};
  \node at (-14.5, 2.5*\linksep-0.5*\eprsep) {$A_{6}$};
  \node at (-14.5, 3.5*\linksep-0.5*\eprsep) {$A_{5}$};
  \node at (-14.5, 4.5*\linksep-0.5*\eprsep) {$A_{4}$};
  \node at (-14.5, 5.5*\linksep-0.5*\eprsep) {$A_{3}$};
  \node at (-14.5, 6.5*\linksep-0.5*\eprsep) {$A_{2}$};
  \node at (-14.5, 7*\linksep-0.5*\eprsep + .75) {$A_{1}$};

  \foreach \i in {1,...,7} {
  \node[font=\scriptsize] (EPR\i) at (1.3, \linksep*\i-.5*\eprsep) {EPR};
  }

  \foreach \j in {1,...,7} {
  \draw[link] (S1\j) -- (S2\j) -- (S3\j);
  \draw[link] (L1\j) -- (L2\j) -- (L3\j);
  \draw[link] (U1\j) -- (U2\j) -- (U3\j);
  \draw[link] (V1\j) -- (V2\j) -- (V3\j);
  }

  \foreach \j in {2,...,7} {
  \draw[link] (S1\j) |- ($(0,-.6) + (L1\j)$) -- (L1\j);
  \draw[link] (S3\j) |- ($(0,-1) + (L3\j)$) -- (L3\j);
  \node[error] at (L1\j) {};
  \node[error] at (S1\j) {};
  \node[gate] at (L2\j) {};
  \node[gate] at (S2\j) {};
  \node[error] at (L3\j) {};
  \node[error] at (S3\j) {};
  }

  \foreach \j in {1,...,6} {
  \draw[link] (U1\j) |- ($(0,.6) + (V1\j)$) -- (V1\j);
  \draw[link] (U3\j) |- ($(0,1) + (V3\j)$) -- (V3\j);
  \node[error] at (U1\j) {};
  \node[error] at (V1\j) {};
  \node[gate] at (U2\j) {};
  \node[gate] at (V2\j) {};
  \node[error] at (U3\j) {};
  \node[error] at (V3\j) {};
  }

  \foreach \j in {1,...,5} {

  \node[empty] (PR\j) at ($(.6*\j,0) + (L32)$) {};
  \node[empty] (QR\j) at ($(.6*\j,\linksep-\linksep*\j) + (L37)$) {};
  \node[adapt] at (PR\j) {};
  \node[adapt] at (QR\j) {};

  \node[empty] (PL\j) at ($(-.6*\j,0) + (S32)$) {};
  \node[empty] (QL\j) at ($(-.6*\j,\linksep-\linksep*\j) + (S37)$) {};
  \node[adapt] at (PL\j) {};
  \node[adapt] at (QL\j) {};

  \node[empty] (MR\j) at ($(.6*\j+5,0) + (U31)$) {};
  \node[empty] (NR\j) at ($(.6*\j+5,\linksep-\linksep*\j) + (U36)$) {};
  \node[adapt] at (MR\j) {};
  \node[adapt] at (NR\j) {};

  \node[empty] (ML\j) at ($(-.6*\j-5,0) + (V31)$) {};
  \node[empty] (NL\j) at ($(-.6*\j-5,\linksep-\linksep*\j) + (V36)$) {};
  \node[adapt] at (ML\j) {};
  \node[adapt] at (NL\j) {};

  \draw[link] (PR\j) -- (QR\j);
  \draw[link] (PL\j) -- (QL\j);
  \draw[link] (MR\j) -- (NR\j);
  \draw[link] (ML\j) -- (NL\j);
  }

  \node[adapt] (PR6) at ($(.6*6,0) + (L32)$) {};
  \node[error] (PR7) at ($(.6*7,0) + (L32)$) {};
  \node[adapt] (XR) at ($(0,-\linksep) + (PR6)$) {};
  \node[error] (XRE) at ($(0,-\linksep) + (PR7)$) {};
  \draw[link] (PR6) -- (XR);
  \draw[link] (PR7) -- (XRE);

  \node[adapt] (MR6) at ($(.6*6+5,0) + (U31)$) {};
  \node[error] (MR7) at ($(.6*7+5,0) + (U31)$) {};
  \node[adapt] (ZR) at ($(0,-\eprsep) + (MR6)$) {};
  \node[error] (ZRE) at ($(0,-\eprsep) + (MR7)$) {};
  \draw[link] (MR6) -- (ZR);
  \draw[link] (MR7) -- (ZRE);

  \node[adapt] (PL6) at ($(-.6*6,0) + (S32)$) {};
  \node[error] (PL7) at ($(-.6*7,0) + (S32)$) {};
  \node[adapt] (XL) at ($(0,-\linksep) + (PL6)$) {};
  \node[error] (XLE) at ($(0,-\linksep) + (PL7)$) {};
  \draw[link] (PL6) -- (XL);
  \draw[link] (PL7) -- (XLE);

  \node[adapt] (ML6) at ($(-.6*6-5,0) + (V31)$) {};
  \node[error] (ML7) at ($(-.6*7-5,0) + (V31)$) {};
  \node[adapt] (ZL) at ($(0,-\eprsep) + (ML6)$) {};
  \node[error] (ZLE) at ($(0,-\eprsep) + (ML7)$) {};
  \draw[link] (ML6) -- (ZL);
  \draw[link] (ML7) -- (ZLE);

  \draw[link] (L32) -- (PR1) -- (PR2) -- (PR3) -- (PR4) -- (PR5) -- (PR6) -- (PR7);
  \draw[link] (S32) -- (PL1) -- (PL2) -- (PL3) -- (PL4) -- (PL5) -- (PL6) -- (PL7);

  \draw[link] (PL7) -| ($(-.6,-.7) + (PL7)$) |- ($(-1,-1.4) + (PR7)$) -|
  ($(.6,-.7) + (PR7)$) |- (PR7); %

  \draw[link] (U31) -- (MR1) -- (MR2) -- (MR3) -- (MR4) -- (MR5) -- (MR6) -- (MR7);
  \draw[link] (V31) -- (ML1) -- (ML2) -- (ML3) -- (ML4) -- (ML5) -- (ML6) -- (ML7);
  \draw[link] (ML7) -| ($(-.6,.7) + (ML7)$) |- ($(-1,1.4) + (MR7)$) -|
  ($(.6,.7) + (MR7)$) |- (MR7); %

  \foreach \j in {1,...,7} {
  \draw[link] (V1\j) -| ($(.6,0) + (S1\j)$) -- (S1\j);
  \draw[link] (U1\j) -| ($(-.6,0) + (L1\j)$) -- (L1\j);
  }

  \foreach \j in {1,...,5} { %
  \draw[link] (QL\j) -| ($(-.6,-.7) + (QL\j)$) |- ($(-1,-1.4) + (QR\j)$) -|
  ($(.6,-.7) + (QR\j)$) |- (QR\j); %
  \draw[link] (NL\j) -| ($(-.6,.7) + (NL\j)$) |- ($(-1,1.4) + (NR\j)$) -|
  ($(.6,.7) + (NR\j)$) |- (NR\j); %
  }

  \foreach[evaluate=\j as \i using int(8-\j)] \j in {1,...,5} {
  \draw[link] (QL\j) -- (S3\i);
  \draw[link] (QR\j) -- (L3\i);
  }

  \foreach[evaluate=\j as \i using int(7-\j)] \j in {1,...,5} {
  \draw[link] (NL\j) -- (V3\i);
  \draw[link] (NR\j) -- (U3\i);
  }

  \foreach[evaluate=\j as \i using int(\j+1)] \j in {1,...,6} {
  \draw[link] (U2\j) -- (L2\i);
  \draw[link] (V2\j) -- (S2\i);
  }

  \node[error] at (S31) {};
  \node[error] at (L31) {};
  \draw[link] (S31) -- (S11.east);
  \draw[link] (L31) -- (L11.west);
  \draw[link] (S31) |- ($(0,-.6) + (L31)$) -- (L31);
  \draw[link] (XLE) |- ($(0,-1) + (XRE)$) -- (XRE);
  \draw[link] (ZLE) |- ($(0,-1.4) + (ZRE)$) -- (ZRE);
  \draw[link] (L31) -- (XR) -- (XRE) -- (ZR) -- (ZRE);
  \draw[link] (S31) -- (XL) -- (XLE) -- (ZL) -- (ZLE);

  \node[empty] (IU7) at ($(-11,0)+(V37)$) {};
  \node[empty] (OU7) at ($(11,0)+(U37)$) {};
  \node[empty] (IU1) at ($(-11,0)+(S31)$) {};
  \node[empty] (OU1) at ($(11,0)+(L31)$) {};
  \draw[link] (IU7) -- (V17.east);
  \draw[link] (OU7) -- (U17.west);
  \draw[link] (IU1) -- (ZLE);
  \draw[link] (OU1) -- (ZRE);
  \end{tikzpicture}
  \caption{The tensor network for the chained entanglement swapping if the straightforward delayed measurement is applied.
  In the figure, the EPR states are in the middle, the red tensors denote the noise effects, the blue tensors represent the Bell meaurements, and the green tensors represent the delayed measurements.}\label{fig:chain-swapping}
  
\end{figure}
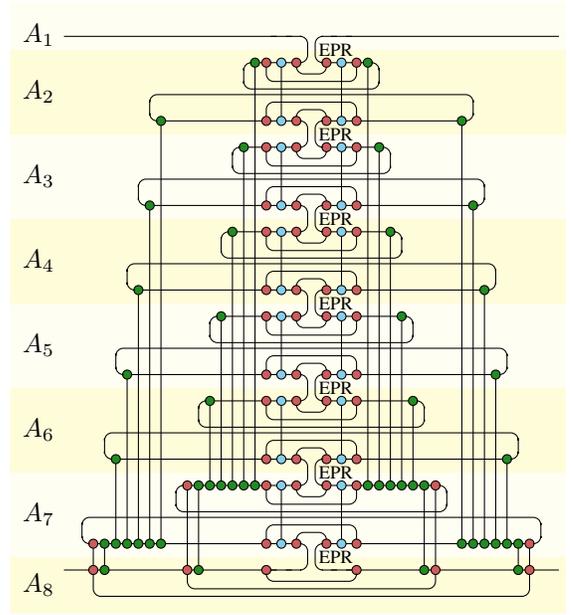

\def\linksep{2}

\begin{figure}[htb]
  \centering 
  \begin{tikzpicture}[scale=.3,
  empty/.style={circle,inner sep=0,outer sep=0,minimum width=1.5mm}, %
  error/.style={draw,circle,fill=royalred,inner sep=0,outer sep=0,minimum width=1.5mm}, %
  gate/.style={draw,circle,fill=royalblue,inner sep=0,outer sep=0,minimum width=1.5mm}, %
  adapt/.style={draw,circle,fill=royalgreen,inner sep=0,outer sep=0,minimum width=1.5mm}, %
  leg/.style={inner sep=0,outer sep=0}, %
  link/.style={rounded corners=1mm}]

  \begin{scope}
  \def\linksep{1}
  \foreach \j in {1,...,7} {
  \draw (-7*1-1, \j * \linksep) -- (0, \j * \linksep);
  }

  \foreach \j in {3,...,7}{
  \node[adapt] (C\j) at (-1*\j, \j*\linksep) {};
  \node[adapt] (T\j) at (-1*\j, 2*\linksep) {};
  \draw[link] (C\j) -- (T\j);
  }
  \node[adapt] (CX) at (-1*2, 2*\linksep) {};
  \node[error] (CE) at (-1*1, 2*\linksep) {};
  \node[adapt] (TX) at (-1*2, \linksep) {};
  \node[error] (TE) at (-1*1, \linksep) {};
  \draw[link] (CX) -- (TX);
  \draw[link] (CE) -- (TE);
  \draw[link] (TE) |- ($(-.7,-.7) + (TE)$);

  \node (arrow) at (1.6,4*\linksep) {$\rightarrow$};
  \end{scope}

  \begin{scope}[xshift=11cm]
  \def\linksep{1}
  \foreach \j in {1,...,7} {
  \draw (-7*1-1, \j * \linksep) -- (0, \j * \linksep);
  }

  \foreach \j in {3,...,7}{
  \node[adapt] (C\j) at (-1*\j, \j*\linksep) {};
  \node[adapt] (T\j) at (-1*\j, \j*\linksep - \linksep) {};
  \draw[link] (C\j) -- (T\j);
  }
  \node[adapt] (CX) at (-1*2, 2*\linksep) {};
  \node[error] (CE) at (-1*1, 2*\linksep) {};
  \node[adapt] (TX) at (-1*2, \linksep) {};
  \node[error] (TE) at (-1*1, \linksep) {};
  \draw[link] (CX) -- (TX);
  \draw[link] (CE) -- (TE);
  \draw[link] (TE) |- ($(-.7,-.7) + (TE)$);
  \end{scope}
  \end{tikzpicture}
  \caption{This figure focuses on the adaption unitary part of \cref{fig:chain-swapping} only.
  When the control flow adaption unitary is chosen appropriately, the corresponding tensor network has lower contraction complexity.}\label{fig:chain-swapping-adapt}
\end{figure}
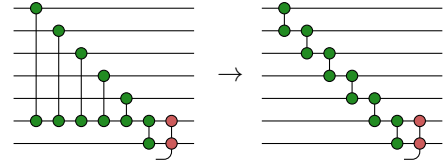

\subsubsection{Entanglement Distillation}

Next, we study the entanglement distillation protocol.
The protocol uses multiple copies of imperfect EPR pairs to ``distill'' an EPR pair with higher quality.
We will first give the protocol of single entanglement distillation~\cite{bennett1996mixed, D_r_2007}, describing how to increase fidelity using two copies of EPR pairs.
Then we will give the protocol of nested entanglement distillation, which extends the protocol into a more general case.

\begin{algorithm}[ht]
  \floatname{algorithm}{Protocol}
  \caption{Single Entanglement Distillation}
  \label{protocol:singledistillation}
  \begin{flushleft}
  \textit{Sender:} Alice, \textit{Receiver:} Bob.\\
  \textit{Input:} $2$ shared EPR pairs with fidelity $F$.
  Each party holds 1 qubit of each pair.
  \\
  \textit{Goal:} Obtain one EPR pair with higher fidelity $F_s>F$, with success probability $p_{s}$.

  \textit{The protocol:}
  \end{flushleft}
  \begin{enumerate}
  \item \textbf{BXOR Test:} Alice and Bob apply a \textsf{CNOT} gate on their own 2
  qubits, and measure their second qubit under the standard basis.
  \item \textbf{Classical Communication:} Alice and Bob inform each other the measurement outcomes.
  \item \textbf{Final Decision:} If Alice and Bob get the same outcomes, the distillation process succeeds.
  Otherwise, if their measurement outcomes are different, the distillation process fails.
  If the process succeeds, they keep the final EPR pair (with higher fidelity $F_{s}$) for further uses.
  If the process fails, they discard the final EPR pair.
  \end{enumerate}
\end{algorithm}

\begin{algorithm}[ht]
  \floatname{algorithm}{Protocol}
  \caption{Nested Entanglement Distillation}\label{protocol:nestedistillation}
  \begin{flushleft}
  \textit{Sender:} Alice, \textit{Receiver:} Bob.

  \textit{Input:} $n=2^k$ shared EPR pairs with fidelity $F_1$.
  Each party holds 1 qubit of each pair.

  \textit{Goal:} Obtain one EPR pair with higher fidelity $F_k>F_1$, with success probability $p_{s}$.  

  \textit{The protocol:}  
  \end{flushleft}
  \begin{enumerate}
  \item \textbf{Recursion:} If $n > 2$, Alice and Bob run the nested distillation protocol twice, with $n/2$ pairs as input each time to obtain 2 EPR pairs.
  Then, Alice and Bob run the single distillation protocol to obtain one EPR pair.

  Else if $n=2$, Alice and Bob directly run the single distillation protocol in Protocol~\ref{protocol:singledistillation}.

  \item \textbf{Final State: }Alice and Bob have one EPR pair with fidelity $F_{k}$, with some success probability $p_s$.

  \item \textbf{Final Decision: } Alice and Bob keep the final EPR pair if and only if all single distillations succeed, and discard the final EPR pair as long as there exists one failure.
  \end{enumerate}
\end{algorithm}
By the distillation protocol, the two parties will use their measurement outcomes to see whether the process succeeds.
Suppose in a round, the measured qubits for the two parties are denoted as $q_1^A, q_2^A, \ldots, q_n^A$ and $q_1^B, q_2^B, \ldots, q_n^B$, with outcomes $m_1^A, m_2^A, \ldots, m_n^A$ and $m_1^B, m_2^B, \ldots, m_n^B$.
And their communication is calculating a function:
\begin{align}
  &\, f(m_1^A, m_2^A, \ldots, m_n^A, m_1^B, m_2^B, \ldots, m_n^B) \nonumber\\
  = &\, (m_1^A\oplus m_1^B)\vee (m_2^A\oplus m_2^B) \vee\ldots\vee(m_n^A\oplus m_n^B).
\end{align}

Here, $\oplus$ denotes Exclusive OR, and $\vee$ denotes OR.
The round of distillation succeeds only if the function output is 0.
Then, by the method, in the simulator, the calculation can be replaced by adding one extra qubit $\textsf R$
initialized to $\ket{0}$, and applying a unitary $U$ such that \begin{align*}
  &\, U \ket{q_1^A,\ldots, q_n^A, q_1^B,\ldots,q_n^B,0} \nonumber\\
  = &\, \ket{q_1^A,\ldots, q_n^A,q_1^B,\ldots,q_n^B,
  f(q_1^A,\ldots, q_n^A,q_1^B,\ldots,q_n^B)},
\end{align*}
and then contracting all other qubits immediately except for $\textsf R$.
This can be achieved by \textsf{CNOT} gates and Toffoli gates.
In this protocol, future actions only depend on whether the process succeeds or fails, and thus the information needed to be recorded only requires an indicator bit.
Here $\textsf R$ acts as the indicator, and the tensor network can immediately simulate whether the process succeeds by simulating the measurement on $\textsf R$.
In this example, the control flow adaption acts like a data compressing process,
which computes the success or failure indicator and passes this information to the next round.
The design of the adaption unitary is actually easy as the straightforward delayed measurement will lead to a solution.
As a result, the control flow will have a tree-like tensor network structure where contractions can be done efficiently.

\subsection{Advantages of Control Flow Adaption}

The advantages of the control flow adaption method are at least two-fold.

First, it enables the exact representation of all the possible measurement branches simultaneously in the tensor network, avoiding the need for Monte Carlo simulations completely.
We take the nested entanglement distillation as an example.
If one uses the Monte Carlo simulation, the simulation is a series of independent experiments with a Boolean-valued outcome (success or fail), and the result follows a binomial distribution.
Therefore, to estimate the final fidelity $F_k$, one needs at least one successful experiment, and the expected number of rounds is $ 1 / p_s$.
To estimate the success probability with precision $\epsilon$, $n$ should satisfies $n > {p(1-p)}/{\epsilon^2}$ by the Chernoff bound.
However, by using our simulator with CFA, one can obtain both the success probability $p_s$ and final fidelity $F_k$ exactly in a single round.

One may question whether a simulation is needed to extract information in this protocol, because existing works~\cite{D_r_2007} have formalized the success probability and resulting fidelity of different entanglement distillation protocols.
We stress the importance of conducting simulations due to the dynamic nature of quantum noise influenced by time delays in classical communication and varying network environments. This variability complicates the direct application of existing theoretical analyses to evaluate the success probability and efficiency of the protocol.

Second, the control flow adaption method reduces the memory usage and the tensor contraction complexity of the underlying tensor network.
In \cref{sec:experiment}, we will demonstrate the advantages of this method through a detailed presentation of some extensive experimental results.

\section{Network Model}\label{sec:netowrk-model}

This section provides an overview of the quantum network model incorporated in the qns-3 simulator.
A quantum network model involves two general classes of components: quantum nodes and quantum channels.
Quantum nodes are entities that manage all equipment capable of processing qubits locally, such as quantum processors, beam splitters, and quantum repeaters.
Quantum channels, such as optical fibers, refer to all communication channels that can transmit quantum information.
While there is still ongoing exploration to determine the proper overall qunamtum network architecture and no definitive consensus has been reached, these two classes of components are universally recognized as essential for the functioning of a quantum network.

\subsection{Quantum Nodes}

Quantum nodes are similar to the end hosts in classical networks, which are connected to other quantum nodes by quantum channels.
In our simulator, the concept of a quantum node is also similar to a node in ns-3.
Each quantum node is equipped with (1) quantum memory devices for storing and manipulating quantum information locally and (2) quantum network devices that link the node to a bunch of quantum channels.
In addition to network hardware, each quantum node is also equipped with network software stacks.
For the classical network stack, we reuse the ns-3 internet stacks and can support any classical network topology.\footnote{In a quantum network, it is also crucial to maintain a classical topology in certain scenarios, such as the CHSH experiment~\cite{clauser1969proposed} where Alice and Bob's classical communication is restricted.}

The model of quantum network protocol stack is still under exploration today.
In this work and the future development of qns-3, we adopt a quantum analog of the classical OSI five-layer model~\cite{kurose2005computer}.
The physical layer is only responsible for transmitting qubits and generating EPR pairs~\cite{dahlberg2019link,panigrahy2022optimal}.
The link layer is responsible for improving the quality of EPR pairs. The most commonly employed techniques are quantum repeaters with entanglement distillation~\cite{QuantumRepeatersPRL} or quantum error correction codes~\cite{QREncoding}.
The network layer is responsible for generating EPR pairs for nodes with long-distance and is loaded with routing algorithms~\cite{zeng2022multi,shi2020concurrent,caleffi2017optimal,farahbakhsh2022opportunistic}.
The transport layer is responsible for delivering qubits to the appropriate application process and transmission control~\cite{yu2021protocols}.
The application layer will run different processes according to the applications implemented on the nodes.
\subsection{Quantum Channels}

A quantum channel connects two quantum nodes and supports the transmission of qubits between the two nodes.
The physical material of quantum channels may be polarization-maintaining optical fibers.
Quantum channels also have the ability to transmit classical information by default.
Quantum channels exhibit inherent noises, potentially leading to qubit errors due to distinct physical attributes associated with each channel.
Hence, specific noise parameters are attributed to each quantum channel.

A general quantum channel could be used to transmit arbitrary qubits.
To realize a reliable transition of qubits through noisy channels, two classes of methods are mostly adopted.
One is the distill-and-teleport approach, which first employs the entanglement distillation protocol to obtain high-fidelity entanglement and then performs quantum teleportation to transmit qubits.
The other is the quantum error correction code (QECC) approach, which encodes the quantum state using a QECC, and directly sends the encoded state through a noisy quantum channel.
Note that the two methods can also be combined, and there are some works~\cite{rijsman2023architectural,muralidharan2016optimal} analyzing how to choose and combine different approaches in suitable scenarios.

Our simulator allows both of them because it supports general quantum operations and quantum channels, which will be further explained in \cref{sec:structure}. As previously mentioned, in the distill-and-teleport protocol, low-fidelity EPR pairs are created and distributed to each party through noisy quantum channels.
After the distribution, parties rely solely on LOCC to complete the protocol. We have integrated the CFA method into the the distill-and-teleport protocol, with its advantages numerically outlined in Section~\ref{sec:experiment}. We also plan to further implement the QECC approach combined with the CFA method in qns-3 in the near future.

\section{Simulator Structure}\label{sec:structure}

\begin{figure*}[h]
  \centering %
  \includegraphics[scale=0.7]{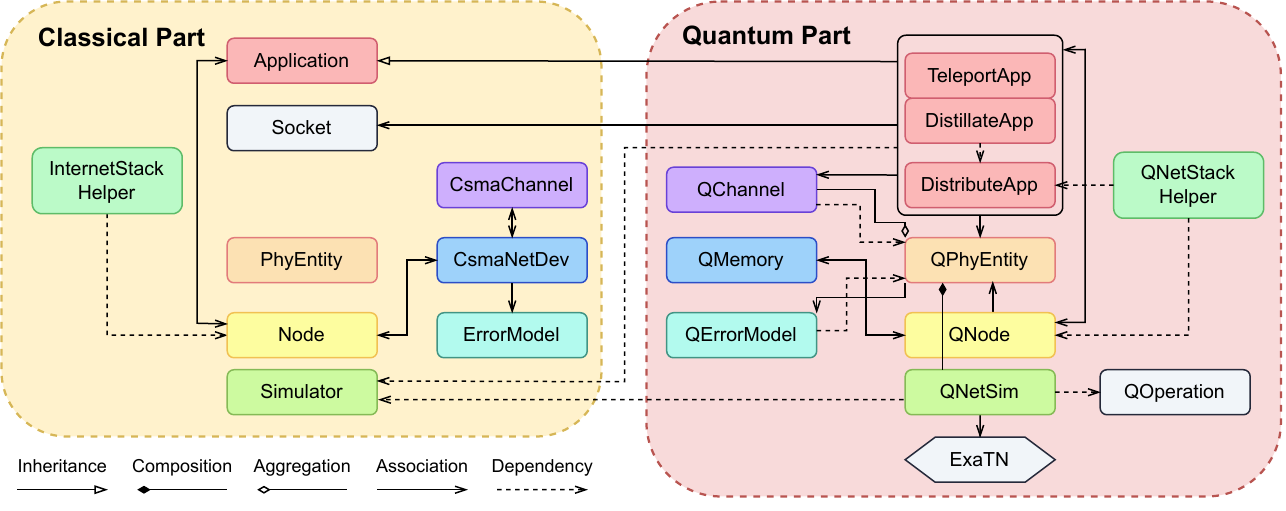}
  \caption{Class diagram of qns-3 in line with the UML standard~\cite{umltutorial}.
  The components are basically split between classical and quantum parts. 
  The left shows the classical network components we mainly use, while the right with the same color shows their counterparts.
  Components with no correspondences are colored grey.
  From a vertical perspective, the architecture begins with underlying simulation engines, followed by network nodes. Atop these, the network stack is situated, with the application layer positioned at the highest level, offering users the flexibility to develop various customized protocols.}\label{fig:structure}
  \end{figure*}
  
In this section, we provide an in-depth examination of our implementation of qns-3. For a more thorough understanding, including a step-by-step guide, we direct readers to our tutorial documentation~\cite{GithubRepo}.

The design adheres to both the traditional OSI layered model of classical networks~\cite{kurose2005computer} and the discrete-event model of ns-3.
It extends the ns-3 framework by introducing a family of quantum modules that encompasses a comprehensive set of quantum network-related features.
Moreover, this extension presents the opportunity to implement our control flow adaption method.

The qns-3 module utilizes tensor networks to represent all qubits, gates and quantum operations internally, and adopts the Exascale Tensor Network (ExaTN)~\cite{lyakh2022exatn} library as the backend.\footnote{ ExaTN is a scalable GPU-accelerated C++ library that can express and process tensor networks on shared as well as distributed memory high-performance computing platforms, including those equipped with GPU accelerators.}
\textsf{QuantumNetworkSimulator} (\textsf{QNetSim} for short) takes full charge of maintaining the quantum circuit represented by exatn:TensorNetwork.
The class encapsulates all direct library calls to ExaTN, providing APIs for creating and initializing qubits, and performing operations and measurements.

\textsf{QuantumNode} (\textsf{QNode}) inherits \textsf{ns3:Node} and has access to the full InternetStack of ns3.
\textsf{QuantumMemory} (\textsf{QMemory}) with configurable \textsf{QuantumErrorModel} (\textsf{QErrorModel}) models the qubit storage in quantum networks and manages the access control of qubit operations.
\textsf{QuantumChannel} (\textsf{QChannel}) models the transmission of quantum information.
\textsf{QMemory} and \textsf{QChannel} are the key interfaces we provide to \textsf{QNode} either directly or indirectly through \textsf{QuantumPhyEntity}.

The following components constitute the quantum network stack:
\textsf{QuantumPhyEntity} (\textsf{QPhyEntity}) is named after \textsf{ns3:PhyEntity} and acts as a quantum counterpart of the classical physical layer.
With the help of \textsf{QNetSim}, it is responsible for managing quantum errors and providing operation APIs to \textsf{QNode}.
\textsf{QuantumNetworkStackHelper} (\textsf{QNetStackHelper}) helps users to install the link layer protocol.
Though not a concern for this work, we mention that it is possible to upgrade the network stack and support other routing and transportation protocols like \textsf{ns3:InternetStack}
\textsf{Helper} does.
As examples of the application and link layer, we implement \textsf{TeleportationApplication} (\textsf{TeleportApp}) and \textsf{DistillationApplication} (\textsf{DistillApp}).
The two applications are implemented with and without the control flow adaption.

The \textsf{QErrorModel} class encapsulates all APIs related to quantum noises.
We emphasize that our simulator supports all kinds of noises, as long as the noise can be represented as a CPTP map.
However, we also define a set of commonly used noise models and seamlessly integrate them into qns-3 to simplify the simulation.
By investigating different causes of quantum noise, we found that four kinds of noises are the most prevalent and frequently considered~\cite{dahlberg2019link, dahlberg2022netqasm, wallnofer2023requsim} for simulating quantum networks: loss error, physical operation noise, quantum channel noise, and time-relevant noise.
\begin{enumerate}
  \item \textit{Loss Error}.
  The loss error denotes the probability of losing a photon during the transition, and each quantum channel is assigned a loss error probability in our simulator.
  \item \textit{Physical Operation Noise}.
  Physical operation noise describes the noise resulting from quantum operations.
  The influence of physical operation noise can be modeled as applying an ideal quantum operation followed by a dephasing error~\cite{coopmans2021netsquid,dahlberg2022netqasm}:
  \begin{align}
  & \mathcal{N}_{\text{dephase}}^f:
  \rho \mapsto f \rho+(1-f) Z \rho Z^\dagger.
  \end{align}
  \item \textit{Quantum Channel Noise.}
  
  When generating noisy EPR pairs, we may focus on the generation fidelity.
  Users can give the initial fidelity of generation $F$, and we simulate the generation process by generating a Werner state $ \rho_{\rm gen} = F \ket{\phi_+}\bra{\phi_+} +
  \frac{1-F}{3} (I-\ket{\phi_+}\bra{\phi_+})$~\cite{lee2000entanglement}.

  \item \textit{Time-relevant Noise}.
  The time-relevant noise describes how qubits decohere over time when kept in the memory~\cite{dahlberg2022netqasm}.
  It can be modeled as ``memory dephasing'' as follows:
  \begin{align*}
  \rho \mapsto (1-\lambda(t)) \rho+\lambda(t)Z\rho Z^\dagger,\\
  {\text {where}}~\lambda(t)=(1-e^{-t/T_{\text{dp}}})/{2}.
  \end{align*}
  Here $T_{\text{dp}}$ is the dephasing time, also the $T_2$ time.
  The numerical value varies from $10~{\rm ms}$ to $1~{\rm s}$ in different physical systems~\cite{wallnofer2023requsim,dahlberg2022netqasm}.
  In our implementation, the \textsf{QErrorModel} module provides built-in support for all kinds of errors mentioned above.
  Quantum time-relevant noise is continuous, but could also be ``discretized'' and simulated in the discrete-event simulator.
  To achieve time-relevant noise using discrete events, we maintain each qubit's \textit{last check time} $t_{\text {last}}$.
  When an event occurs, the simulator will apply appropriate dephasing noise to all the qubits involved in this event with the parameter $t=t_{\text{current}}-t_{\text{last}}$ before simulating the event.
\end{enumerate}


\section{Experiment Results}\label{sec:experiment}

\subsection{Experiment Setup}

Our tests are performed on a Linux workstation using 16 vCPUs at 3.30GHz (Intel(R) Xeon(R) Platinum 8369HB) and 60 GiB RAM, sufficient to demonstrate the advantages of employing CFA as well as other computation needed for comparison.
All codes are open-sourced in~\cite{GithubRepo}.


\subsubsection{Metrics}

Metrics used in the following evaluations are runtime and memory usage for a simulation round, which indicate the time and space cost of the simulator.
Runtime is evaluated in terms of the wall-clock runtime.
Memory usage is evaluated by the peak memory usage of the process.
Those two metrics are used to evaluate the performance of different CFA methods,
including the trivial one without any adaptation.

\subsubsection{Benchmarks}

We use as benchmarks two quantum network protocols including the chained entanglement swapping and the nested entanglement distillation, evaluated at scales involving up to thousands of qubits.
In the chained entanglement swapping protocol, all nodes are arranged in a line topology.
Every node performs an entanglement swapping with its successor, passing the entanglement down to the last node.
As for the nested entanglement distillation protocol, every nested distillation process is a composite of smaller-scale, isomorphic sub-processes.
Since each distillation takes two EPR pairs and gives one, the number of EPR pairs initially prepared is set to powers of two.

\subsubsection{Comparasion}

We compare the performance of qns-3 with NetSquid~\cite{dahlberg2022netqasm},
which we believe is the best reference for the performance of a quantum network simulator.
Because NetSquid faithfully simulates all events in a network, we combine it with the Monte Carlo method to obtain an average result when simulating the benchmark protocols.
In the entanglement swapping, we set the number of Monte Carlo rounds to $10^4$,
achieving approximately $0.01$ precision.
In the nested distillation, we set the number of Monte Carlo rounds to $200$, achieving approximately $0.1$ precision.
We also compare qns-3 with NetSquid in terms of runtime and memory usage.

\subsection{Runtime and Memory Usage}

The following experiments are to demonstrate the separation in time and space performance of different CFA methods using the two benchmark protocols.

\subsubsection{Chained Entanglement Swapping}

\begin{figure}[h!]
  \centering \begin{subfigure}[b]{0.235\textwidth}
  \includegraphics[width=\textwidth]{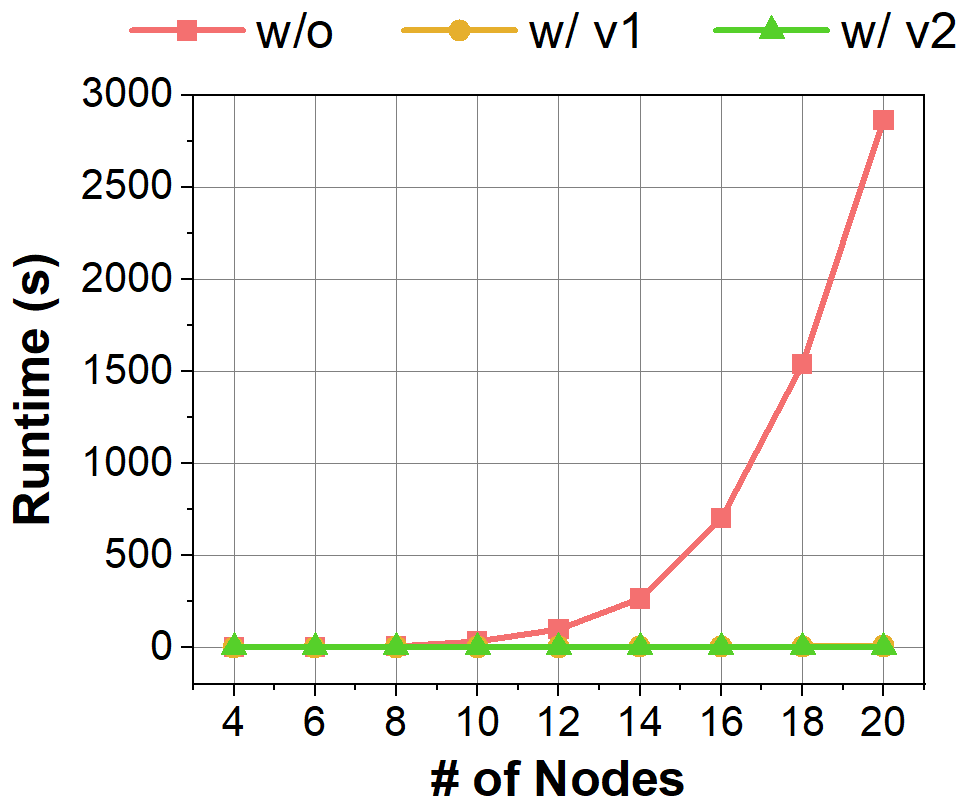}
  \caption{Runtime.}
  \label{fig:swap-adapt-time}
  \end{subfigure}
  \hfill \begin{subfigure}[b]{0.23\textwidth}
  \includegraphics[width=\textwidth]{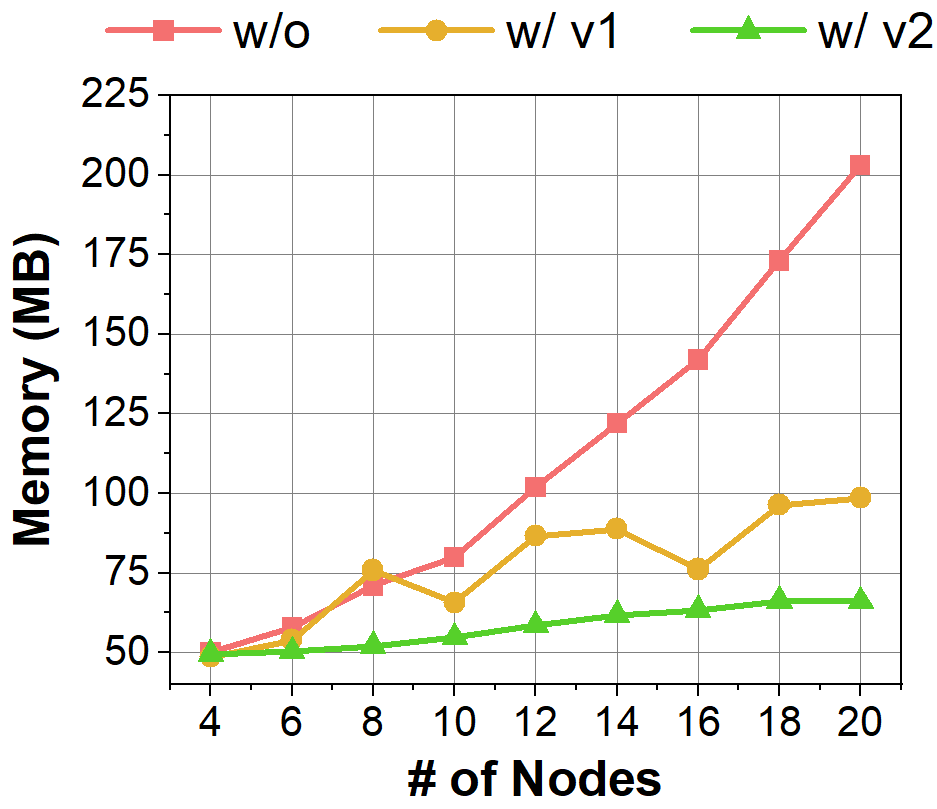}
  \caption{Memory.}
  \label{fig:swap-adapt-space}
  \end{subfigure}
  \caption{ A demonstration of the advantage of using adaptation when simulating the chained entanglement swapping protocol in qns-3.
  Time and memory costs are plotted as a function of the number of nodes in the chain.
  The pink curve grows exponentially, with a direct yet naive implementation of the protocol.
  The green curve with the second adaptation is linear.}\label{fig:swap-adapt}
\end{figure}

\begin{figure}[h!]
  \centering \begin{subfigure}[b]{0.23\textwidth}
  \includegraphics[width=\textwidth]{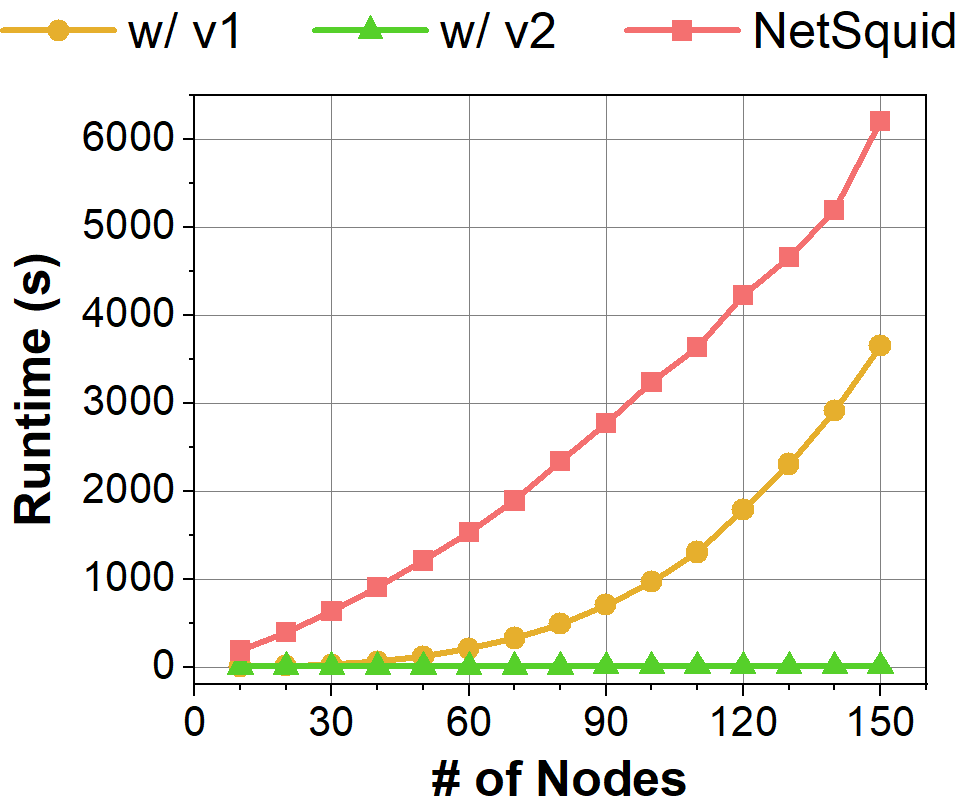}
  \caption{Runtime.}
  \label{fig:swap-netsquid-time}
  \end{subfigure}
  \hfill \begin{subfigure}[b]{0.23\textwidth}
  \includegraphics[width=\textwidth]{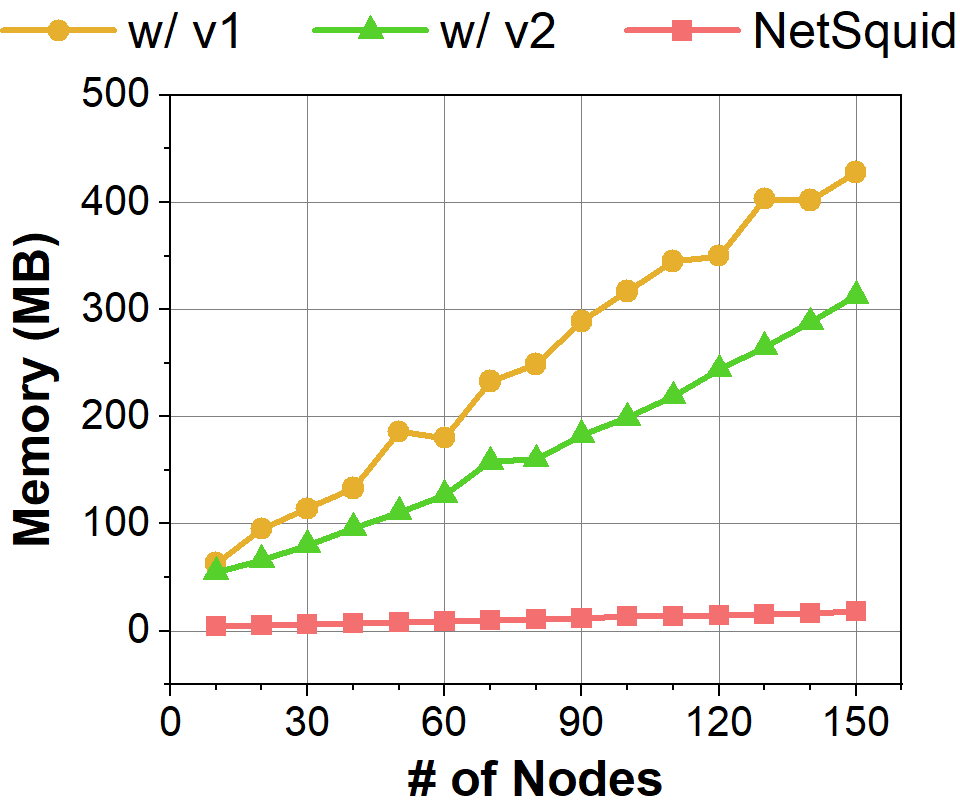}
  \caption{Memory.}\label{fig:swap-netsquid-space}
  \end{subfigure}
  \caption{A demonstration of the advantage of qns-3 when simulating the chained entanglement swapping protocol.
  Compared with NetSquid, qns-3 achieves a much shorter run time when the number of nodes increases.
  However, NetSquid uses less memory, for it only faithfully simulates the protocol and reveals information about one branch each time.
  Time and memory costs are functions of the number of nodes in the chain.
  The figure also demonstrates the advantage of designing a better adaptation method in large-scale chained entanglement swapping protocols in qns-3.
  The yellow curve uses the first version of adaptation, while the green curve uses the second.}\label{fig:swap-netsquid}
  \end{figure}

The simulation costs of the chained entanglement swapping protocol are shown in Fig.~\ref{fig:swap-adapt}, and the comparison with NetSquid is shown in Fig.~\ref{fig:swap-netsquid}.
For each CFA method, we vary the length of the chain and test its runtime and memory usage.
The result shows that with CFA, the simulation cost is highly reduced.

The advantage of the CFA method is already evident as illustrated in \cref{fig:swap-adapt}.
Notice that in \cref{fig:swap-adapt-time}, the number of qubits is so small that the runtime curves of the two versions of adaptations (labeled by yellow circles and green triangles) are almost indistinguishable.
Yet, as discussed in \cref{sec:method}, the first method also has exponential complexity, and this can be seen in \cref{fig:swap-netsquid}, by scaling the number of nodes in the chain to about $150$.

\subsubsection{Nested Entanglement Distillation}

\begin{figure}[ht]
  \centering \begin{subfigure}[b]{0.235\textwidth}
  \includegraphics[width=\textwidth]{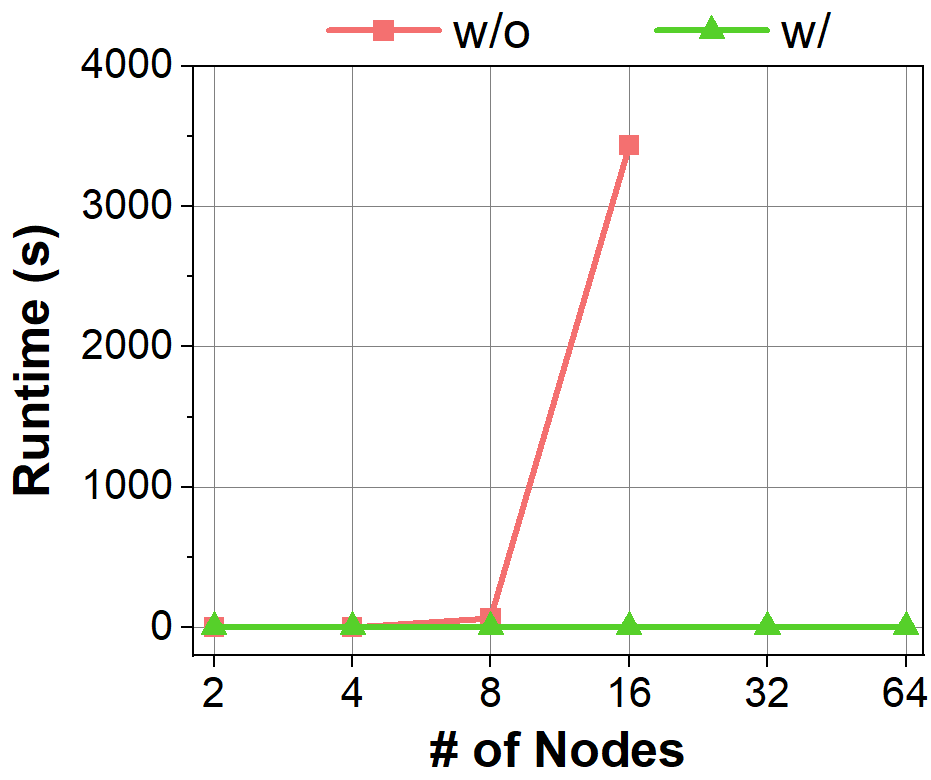}
  \caption{Runtime.}\label{fig:dist-adapt-time}
  \end{subfigure}
  \hfill \begin{subfigure}[b]{0.23\textwidth}
  \includegraphics[width=\textwidth]{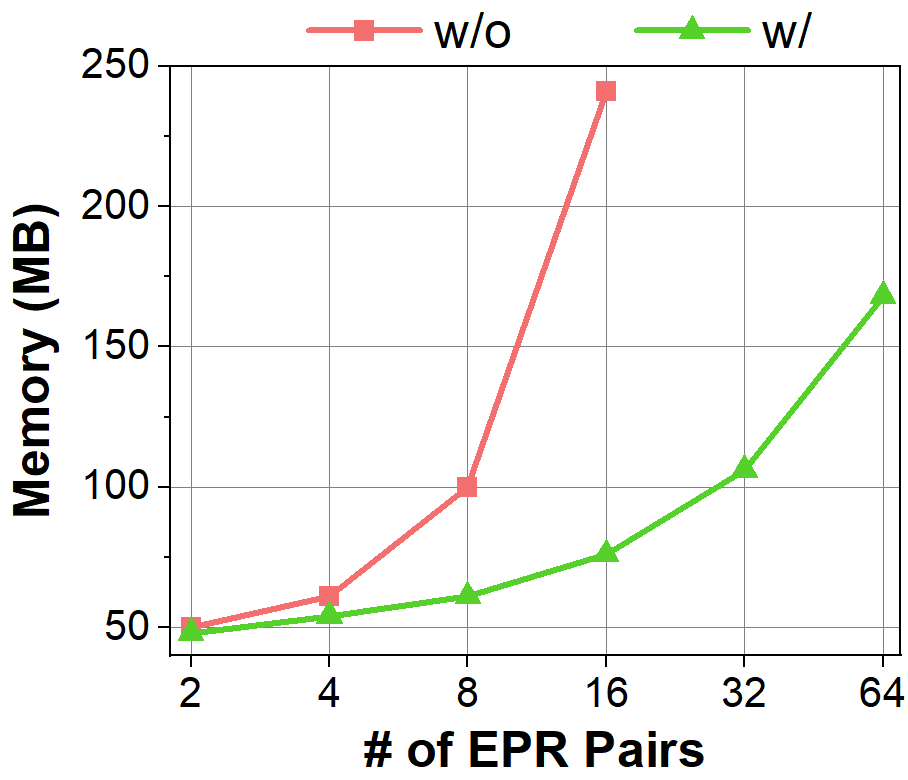}
  \caption{Memory.}\label{fig:dist-adapt-space}
  \end{subfigure}
  \caption{Simulation of the nested entanglement distillation protocol and scaling the number of EPR pairs, to demonstrate the advantage of using adaptation.
  The pink curve, indicating a naive implementation, grows exponentially.
  It is truncated due to the two-hour time-out at 32 pairs.
  The green curve with a proper adaptation is linear.}\label{fig:dist-adapt}
\end{figure}

For the nested distillation protocol, we vary the number of initially prepared EPR pairs and compare the time and space cost in \cref{fig:dist-adapt}.
CFA is able to reduce the cost from exponential to linear, as it trims down the maximum cut of the tensor network from linear to constant, which in turn compresses the complexity of tensor contraction the most time-consuming part in simulation.

\begin{figure}[h!]
  \centering \begin{subfigure}[b]{0.23\textwidth}
  \includegraphics[width=\textwidth]{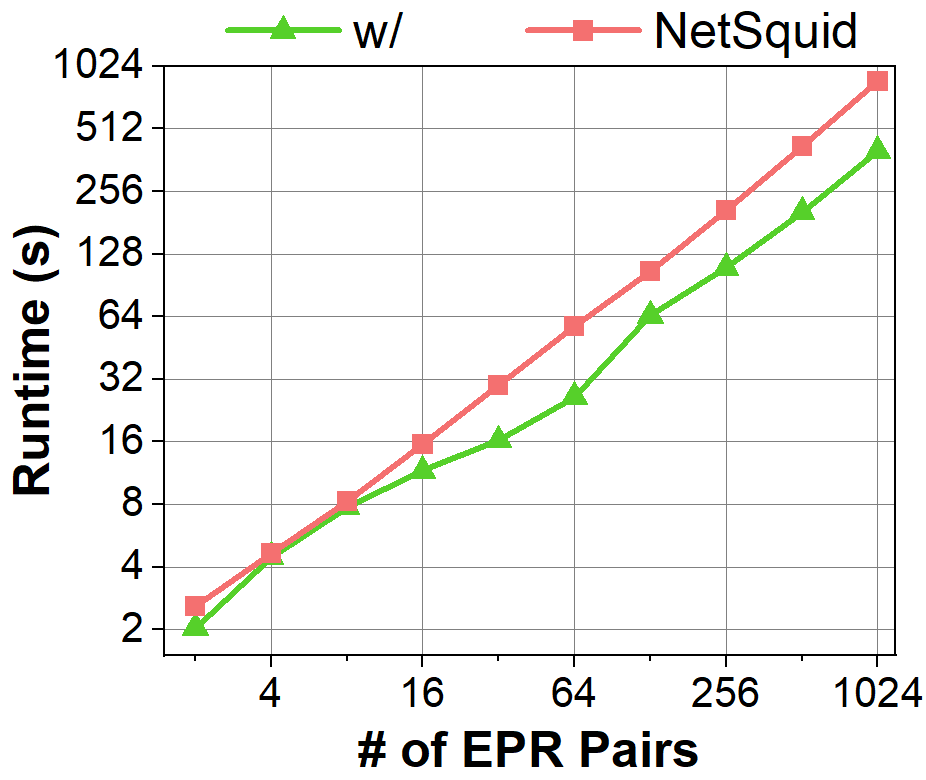}
  \caption{Runtime.}\label{fig:dist-ns-time}
  \end{subfigure}
  \hfill \begin{subfigure}[b]{0.238\textwidth}
  \includegraphics[width=\textwidth]{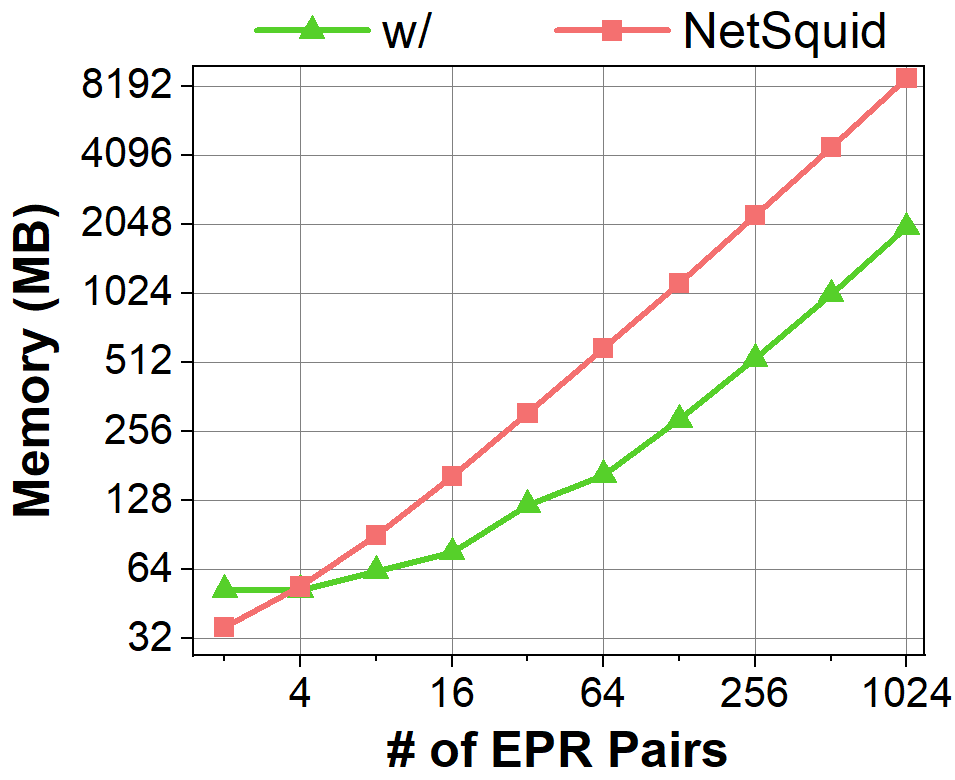}
  \caption{Memory.}\label{fig:dist-ns-space}
  \end{subfigure}
  \caption{A demonstration of the advantage of qns-3 when simulating the the nested distillation protocol.
  Compared with NetSquid, qns-3 achieves shorter run time when the number of EPR pairs increases, and also costs less memory.
  And qns-3 can reveal the precise success probability and final fidelity,
  while the precision of NetSquid is approximately $0.1$, and the cost will increase if higher precision is required.
  This figure also demonstrates that the runtime of qns-3 scales linearly with the number of qubits involved.}\label{fig:dist-ns}
\end{figure}

The experiments show that the simulator performs much better with CFA\@.
The result shown in \cref{fig:dist-ns-time} test demonstrates more directly that the simulator with CFA is scalable, i.e., fast enough that it scales linearly with the number of EPR pairs.

\section{Conclusion}\label{sec:conclusion}

In this paper, we present the control flow adaptation method for quantum network simulation, and introduce the qns-3 platform, the first general quantum network simulator specifically designed for ns-3.
Our method efficiently manages classical control of a protocol within the tensor network data structure, enabling efficient simulations of diverse quantum network protocols.
We strongly believe that both the control flow adaptation method and the qns-3 platform hold significant potential for widespread application in future research on quantum network protocols.

\section*{Acknowledgement}\label{sec:ack}

The work is supported by National Key Research and Development Program of China (Grant No.\ 2023YFA1009403), National Natural Science Foundation of China (Grant No.\ 12347104), Beijing Natural Science Foundation (Grant No.\ Z220002), and a startup funding from Tsinghua University.
Zhengfeng Ji is the corresponding author.

\bibliographystyle{plain}
\bibliography{reference}

\end{document}